\documentclass[12pt]{article}
\usepackage{amsmath}
\renewcommand{\a}{\alpha}
\renewcommand{\b}{\beta}

\begin{document}
\setlength{\baselineskip}{16pt}

\begin{center}
{\large\bf A Note on the Intermediate Region}\smallskip\\
{\large\bf in  Turbulent Boundary Layers}

\vspace{1 truein}

{\sc G. I. Barenblatt,${}^1$ \ A. J. Chorin${}^1$ and V. M.
Prostokishin${}^2$}

\vspace{.33 truein}
${}^1$Department of Mathematics and\\
Lawrence Berkeley National Laboratory \\
University of California\\
Berkeley, California 94720

\bigskip\bigskip

${}^2$P. P. Shirshov Institute of Oceanology\\Russian Academy of
Sciences\\
  36, Nakhimov Prospect\\
  Moscow 117218 Russia
\end{center}

\bigskip\bigskip\bigskip
\begin{center}{\bf Abstract.}\end{center}
We demonstrate that the processing of the experimental data for the average
velocity profiles obtained by J.~M. \"Osterlund
(www.mesh.kth.se/$\sim$jens/zpg/)  presented in [${}^1$] was
incorrect.  Properly processed these data lead to the opposite
conclusion:  they confirm the Reynolds-number-dependent scaling
law and disprove the conclusion that the flow in the
intermediate (`overlap') region is Reynolds-number-independent.

\newpage
In a recent issue of the Physics of Fluids \"Osterlund et
al.~[${}^1$] presented results of the processing of experimental data for the
average velocity field in the zero-pressure-gradient turbulent boundary
layer, published by the first author of [${}^1$] on the Internet \linebreak
({\tt http://www.mesh.kth.x/$\sim$jens/zpg/}\,).  The conclusion of [${}^1$]
is very definite (p.~1):  `Contrary to the conclusions of some earlier
publications, careful analysis of the data reveals no significant Reynolds
number dependence for the parameters describing the overlap region using the
classical logarithmic relation'.

In the present note we show that the processing of the experimental data in
[${}^1$] is incorrect.  A correct processing is performed, and the result
is the opposite:  there is significant Reynolds number dependence for the
parameters describing the intermediate region and the scaling (power) law is
valid.  We also show where the authors of [${}^1$] went wrong.

\bigskip
{\bf The correct processing of the \"Osterlund data.}
If a scaling (power) law relates the dimensionless velocity ${\bar U}^+$ and
dimensionless distance from the wall $y^+$ (in the notations of [${}^1$]):

\medskip
\noindent
(I)\hfill ${\bar U}^+ = A(y^+)^{\a}$\hfill $(1)$

\medskip
\noindent
then in the coordinates $\mbox{lg } y^+$, $\mbox{lg } {\bar U}^+$ the
experimental points should lie within experimental accuracy along a straight
line:  $\mbox{lg } {\bar U}^+ = \mbox{lg } A + \a \mbox{ lg } y^+$. 
Therefore our first step was to plot the data from all 70 runs available on
the Internet in these coordinates.  All 70 runs corresponding to different
$Re_{\theta}$ yield the same pattern:  in the intermediate region
between the viscous sublayer and free stream the average velocity
distribution consists of two straight lines.  Three examples are presented
in Figure~1 and the Table, all the remaining ones are similar and can be
found in our detailed report [${}^2$].  Thus the intermediate structure
(between the viscous sublayer and free stream) consists of two self-similar
layers:  the scaling law (I) is valid for the layer adjacent to the viscous
sublayer, and the scaling law

\medskip
\noindent
(II)\hfill ${\bar U}^+ = B(y^+)^{\b}$\hfill $(2)$

\medskip
\noindent
holds in the layer adjacent to the free stream.  This means, in particular,
that {\em some} scaling law with Reynolds-number-dependent coefficients is
valid in the region (I).  The coefficients $A$ and $\a$, constant for every
run, depend on the Reynolds number which is different for different runs, and
their variation is substantial (see the Table).

Once this is established, we investigate whether this scaling law can be
represented in the form
$$
{\bar U}^+ = \left( \frac {1}{\sqrt{3}} \ln Re + \frac {5}{2}
\right) (y^+)^{\frac {3}{2 \ln  Re}} \eqno(3)
$$
obtained by us earlier for flow in pipes.  In the case of pipe flows $Re$ was
${\bar u}d/\nu$, where ${\bar u}$ is the mean velocity (bulk flux divided by
cross-section area), and $d$ is the diameter. But what is $Re$ for the
boundary layer?

The effective Reynolds number $Re$ should have the form $Re =
U\Lambda/\nu$, where $U$ is the free stream velocity, $\nu$ the kinematic
viscosity, and $\Lambda$ a length scale which we cannot {\em \`a priori}
identify with the momentum thickness $\theta$, as there is no rationale for
such identification.  So, the {\em basic question is whether one can find for
each run a length scale $\Lambda$ so that the scaling law $(3)$ will be
valid for the mean velocity distribution in the first intermediate region.} 
If this scale exists then the law $(3)$ is not specific to flows in pipes but
may be a general law for wall-bounded shear flows at large Reynolds numbers.

To answer this question we took the values $A$ and $\a$ for each run,
obtained by standard statistical processing of the experimental data in the
first intermediate scaling region, and then calculated two values $\mbox{ln
} Re_1$ and $\mbox{ln } Re_2$ by solving two equations suggested by the law
$(3)$,
$$
\frac {1}{\sqrt{3}} \mbox{ln } Re_1 + \frac {5}{2} = A,\quad  \frac {3}{2 
\ln Re_2} = \a. \eqno(4)
$$
If the values $\mbox{ln } Re_1$ and $\mbox{ln } Re_2$ obtained by solving
these two {\em different} equations $(4)$ coincide within experimental
accuracy, then the unique length scale $\Lambda$ can be determined so that
the experimental scaling law in the region $(1)$ coincides with the law
$(3)$.  Indeed these values {\em are} close --- for all $Re_{\theta} >
10,000$, the difference $\Delta =(\mbox{ln } Re_2 - \mbox{ln } Re_1)/\ln Re$
does not exceed 3\%, see the table for the examples in Figure~1 and in
[${}^2$] for all runs.  This allows one to introduce, for large Reynolds
numbers, an effective Reynolds number $Re$, for example by the relation
$$
\mbox{ln } Re_1 = \frac {1}{2} (\mbox{ln } Re_1 + \mbox{ln }
Re_2), 
\mbox{ or } Re = \sqrt{Re_1Re_2}, \eqno(5)
$$
the geometric mean of $Re_1$ and $Re_2$.  This Reynolds number defines the
effective length scale $\Lambda$, which plays for boundary layer flow the
same role as the pipe diameter for flow in pipes.  Remember that the momentum
thickness is calculated by integration of the velocity profile obtained
experimentally:  the calculation of the length scale on the basis of the
measured velocity profile is not more complicated.  Furthermore, the scaling
law $(3)$ can be reduced to a universal form
$$
\psi = \frac {1}{\a} \ln \left( \frac {2\a{\bar U}^+}{\sqrt{3} +
5\a} \right) = \ln y^+ \eqno(6)
$$
where $\a = \frac {3}{2 \ln Re}$.  This formula gives another way to
check the applicability of the Reynolds-number-dependent scaling law $(3)$
in the intermediate region $(1)$.  Indeed, according to $(6)$, in the
coordinates $\mbox{ln } y^+,\psi$, all experimental points should collapse
onto the bisectrix of the first quadrant.  Figure~2 shows that all data for
large Reynolds numbers $(Re_{\theta} > 15,000$, 24~runs) presented on the
Internet collapse onto the bisectrix with accuracy sufficient to give an
additional confirmation to the Reynolds-number-dependent scaling law $(3)$. 
For lesser values of $Re_{\theta}$ a small but systematic parallel shift is
observed (see [${}^2$]).  One possible reason is that in these cases the
choice of
$Re$ according to $(5)$ may be insufficient because in particular at small
Reynolds numbers the higher terms of the expansion of the coefficients of
the scaling law over
$1/\mbox{ln } Re$ could have some influence (see the paper by Radhakrishnan
Srinivasan [${}^3$]); another possibility is that the relation that was to
be proved was used as fact in the calculation of the skin friction in
equation (6) of [${}^1$].

\bigskip\noindent
{\bf Why is the data processing in [${}^1$] incorrect?} \
The main argument against the power law used by the authors of [${}^1$] is
the following.  They introduce the ``diagnostic function'' (p.3, right)
$$
\Gamma = \frac {y^+}{{\bar U}^+} \ \frac {d{\bar U}^+}{dy^+}. \eqno(7)
$$
Their statement, ``The function $\Gamma$ should be a constant in a region
governed by a power law'' is correct for {\em a fixed Reynolds number}. 
However, this is not true for the `{\em diagnostic function averaged for
KTH data}', which is shown in their Figure~6.

We invite the reader to look at Figure~1 (the situation with all other runs
is the same, see our report [${}^2$]).  It is clear that for each run
$\Gamma$, which is equal to $d(\lg{\bar U}^+)/d\lg(y^+)$, is
a constant---look at the straight lines in the first intermediate region! 
However, (see the Table) {\em this constant is different for different runs
because the slope of the straight lines is
$Re$-dependent!}  Indeed, the slope in the first region decays with growing
Reynolds number.  It is clear why $\Gamma$ obtained by the authors based on
averages for KTH data, is decreasing:  the runs with larger Reynolds number
and smaller slopes contribute more at larger $y^+$.

Due to this incorrect procedure the authors were unable to recognize the
Reynolds-number-dependent power law in their rather {\em rich} database.

Their determination of the constants of the logarithmic law gives an
additional illustration of the incorrectness of their procedure.  They do it
in two ways; it is enough to mention what they call the traditional
procedure (p.~3, right).  They take the data representation in the
traditional $\mbox{lg } y^+,{\bar U}^+$ plane, and calculate the constant
$\kappa$ in the logarithmic law
$$
{\bar U}^+ = \frac {1}{\kappa} \ln(y^+) + B \eqno(8)
$$
`by fitting a log-law relation for each profile using the following
traditional limits of the fit:  $M_1 = 50$, and $M_0 = 0.15$'.  The result
of their fit is shown in Figure~3 (their Figure~5) which is far
from convincing.

Apparently embarrassed by this result, the authors extended the upper
boundary of the viscous sublayer to $M_1 = 200$ (the traditional value is
30-70), and obtained, again, with huge scatter, the value $\kappa = 0.38$. 
For the constant $B$ they obtained the value $4.1$.  Both values are the
lowest contenders among the values available in the literature (Nikuradze,
$\kappa = 0.417$, $B\!=\!5.89$; Monin and Yaglom, $\kappa = 0.40$,
$B\!=\!5.1$; Schlichting, $\kappa = 0.40$, $B\!=\!5.5$).  However, if the law
$(8)$ is universal, the values of the constants should be identical for all
high quality experiments!

Finally, according to the authors of [${}^1$], the logarithmic layer extends
only over 1/6 of the boundary layer thickness.  In fact, a better
representation is provided by the power law in the first region. The upper
boundary of this first region is always higher than the upper boundary of the
logarithmic region presented in [${}^1$].

In conclusion, careful analysis of \"Osterlund's data, contrary to the claim
in [${}^1$], reveals a significant Reynolds number dependence for the
parameters describing the `overlap' region and confirms the
Reynolds-number-dependent scaling law.

\bigskip{\bf Acknowledgments.} This work was supported in part by the
Applied Mathematics subprogram of the U.S.~Department of Energy under
contract DE--AC03--76--SF00098, and in part by the National Science
Foundation under grants DMS\,94--16431 and DMS\,97--32710.

\bigskip\bigskip
\begin{center}{\Large\bf References}\end{center}

\begin{enumerate}
\item J. M. \"Osterlund, A. V. Johansson, H. M. Nagib, and M. H.
Hites, ``A note on the overlap region in turbulent boundary layers'',
{\em Phys.~Fluids} {\bf 12}, No.~1, pp.~1--4 (2000).

\item G. I. Barenblatt, A. J. Chorin, and V. M. Prostokishin,
``Analysis of experimental investigations of self-similar intermediate
structures in zero-pressure-gradient boundary layers at large Reynolds
numbers'', UC Berkeley Math Dept Report PAM-777, January~2000.

\item Radhakrishnan
Srinivasan, ``The importance of higher-order
effects in the Barenblatt--Chorin theory of wall-bounded fully developed
turbulent shear flows'', {\em Phys. Fluids} {\bf 10}, No.~4, pp.~1037--1039
(1998).

\end{enumerate}

\bigskip
\begin{center}{\Large\bf Table}\bigskip

\begin{tabular}{ccccccccc}
$Re_{\theta}$ & $\a$ & A & $\b$ & B & $\ln(Re_1)$ & $\ln(Re_2)$ &
$\ln(Re)$ & $\Delta$,\% \medskip\\
2,532 & 0.157 & 7.84 & 0.226 & 5.32 & 9.24 & 9.57 & 9.4 & 3.4\\
14,207& 0.132& 9.01 & 0.191 & 5.87 & 11.28 & 11.39 & 11.33 & 1.0\\
26,612 & 0.120& 9.74& 0.177& 6.24& 12.54 & 12.48& 12.51 & 0.5
\end{tabular}
\end{center}

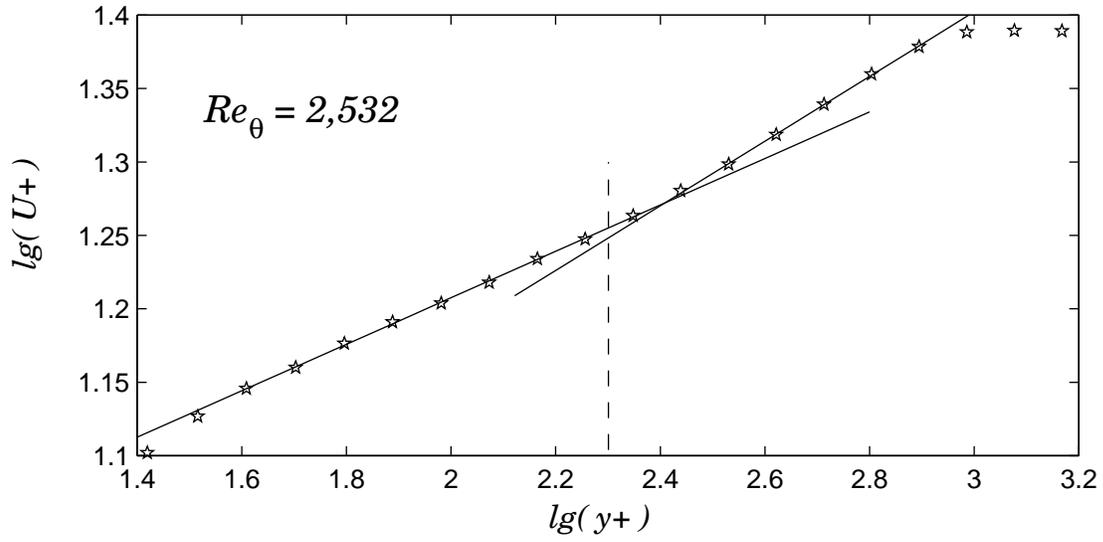
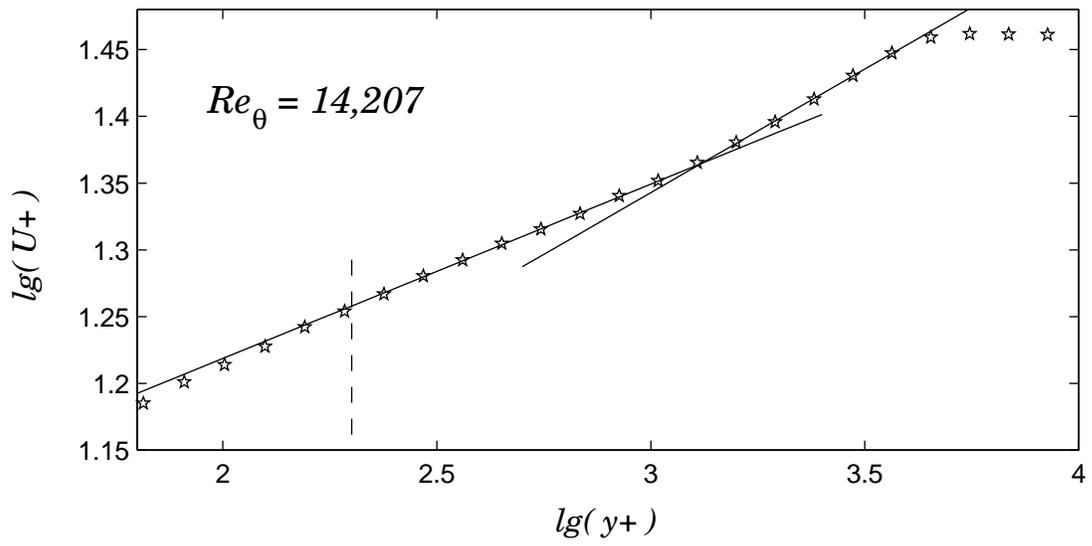
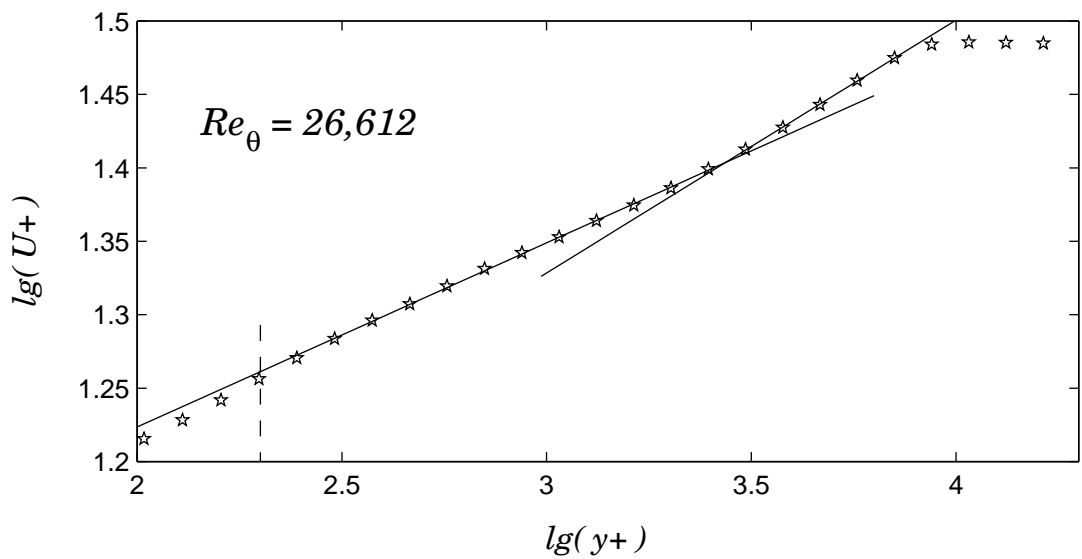
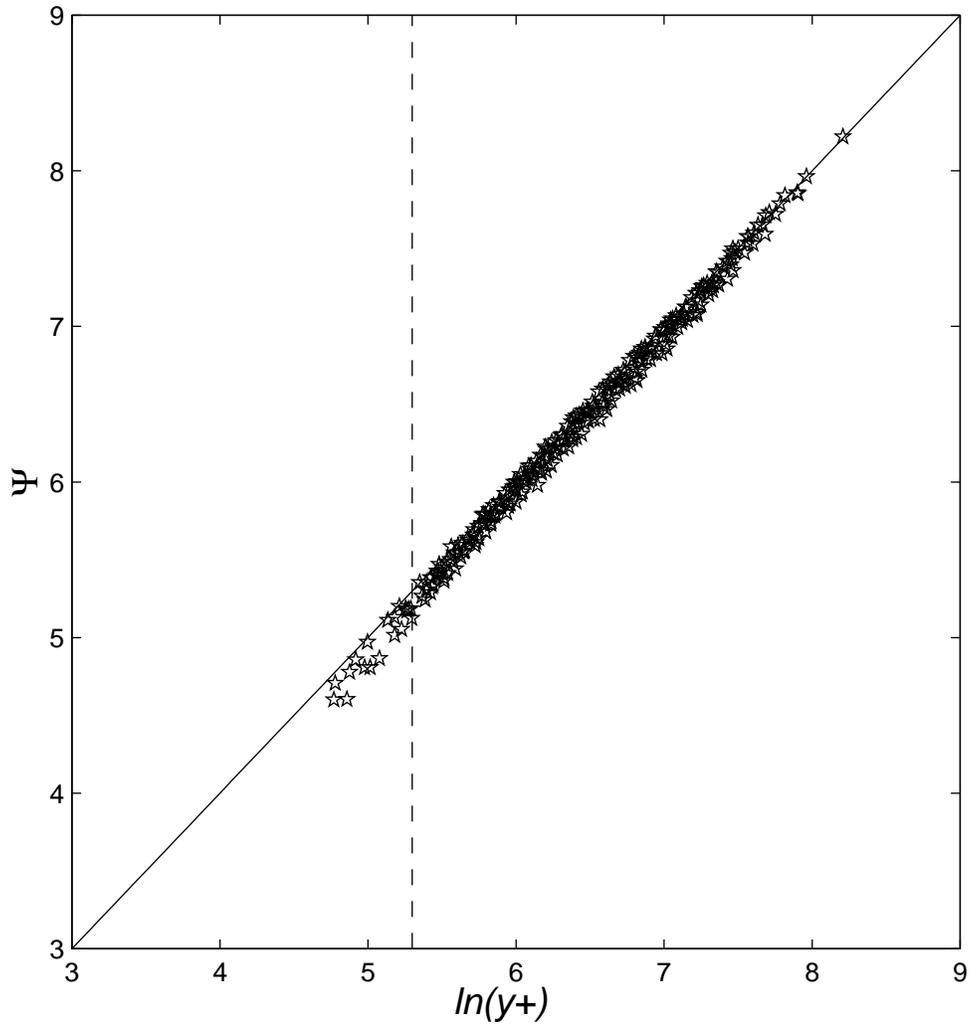
\end{document}